\documentclass[prl,superscriptaddress,twocolumn,preprintnumbers]{revtex4-1}

\usepackage{epsfig}
\usepackage{graphicx}
\usepackage{color}
\usepackage[normalem]{ulem}
\usepackage[breaklinks, plainpages=false, colorlinks=true, anchorcolor=cyan, linkcolor=red, citecolor=cyan, urlcolor=magenta, bookmarks=false]{hyperref}

\usepackage[caption=false]{subfig}

\begin{document}
\renewcommand{\thefigure}{\arabic{figure}}
\setcounter{figure}{0}

 \def\I{{\rm i}}
 \def\E{{\rm e}}
 \def\D{{\rm d}}

\bibliographystyle{apsrev}

\title{Constraining alternative theories of gravity using pulsar timing arrays}

\author{Neil J. Cornish}
\affiliation{eXtreme Gravity Institute, Department of Physics, Montana State University, Bozeman, Montana 59717, USA}

\author{Logan O'Beirne}
\affiliation{eXtreme Gravity Institute, Department of Physics, Montana State University, Bozeman, Montana 59717, USA}

\author{Stephen R. Taylor }
\affiliation{Theoretical AstroPhysics Including Relativity (TAPIR), MC 350-17, California Institute of Technology, Pasadena, California 91125, USA}

\author{Nicol\'{a}s Yunes}
\affiliation{eXtreme Gravity Institute, Department of Physics, Montana State University, Bozeman, Montana 59717, USA}

\begin{abstract}
The opening of the gravitational wave window by ground-based laser interferometers has made possible many new tests of gravity, including the first constraints on polarization.
It is hoped that within the next decade pulsar timing will extend the window by making the first detections in the nano-Hertz frequency regime. Pulsar timing offers several advantages over ground-based interferometers for constraining the polarization of gravitational waves due to the many projections of the polarization pattern provided by the different lines of sight to the pulsars, and the enhanced response to longitudinal polarizations. Here we show that existing results from pulsar timing arrays can be used to place stringent limits on the energy density of longitudinal stochastic gravitational waves. Paradoxically however, we find that longitudinal modes will be very difficult to detect due to the large variance in the pulsar-pulsar correlation patterns for these modes. Existing upper limits on the power spectrum of pulsar timing residuals imply that the amplitude of vector longitudinal and scalar longitudinal modes at frequencies of 1/year are constrained: ${\cal A}_{\rm VL} < 4.1\times 10^{-16}$ and
${\cal A}_{\rm SL} < 3.7\times 10^{-17}$, while the bounds on the energy density for a scale invariant cosmological background are: $\Omega_{\rm VL}h^2  < 3.5  \times 10^{-11}$ and
$\Omega_{\rm SL}h^2  < 3.2  \times 10^{-13}$.
\end{abstract}

\maketitle

The detection of gravitational waves from merging black hole binaries~\cite{Abbott:2016blz,Abbott:2016nmj, Abbott:2017vtc, Abbott:2017gyy, Abbott:2017oio} and neutron stars~\cite{TheLIGOScientific:2017qsa, Monitor:2017mdv, GBM:2017lvd} by the LIGO/Virgo collaboration has made possible many fundamental tests of gravity~\cite{Abbott:2017vtc, TheLIGOScientific:2016src,TheLIGOScientific:2016pea,Yunes:2016jcc}, including the first studies of the polarization content of the waves~\cite{Abbott:2017oio}. Alternatives to Einstein's theory of gravity generically predict the presence of scalar and vector polarization states, in addition to the usual tensor modes~\cite{Will:2005va, lrr-2013-9, Chatziioannou:2012rf}. Pulsars are also a tremendously valuable tool for probing these strong gravity effects. Pulsar timing observations of binary systems have been used to constrain the fraction of the emitted energy that goes into scalar and vector polarization states~\cite{Taylor:1989sw, Stairs:2002cw, Kramer:2006nb}. The LIGO/Virgo collaboration have produced upper limits on the emission of non-tensorial gravitational waves from isolated pulsars~\cite{Abbott:2017tlp}. These results consider pulsars as {\em sources} of gravitational waves. Here we derive new limits on alternative theories of gravity by considering an array of milli-second pulsars as a gravitational wave {\em detector} \citep{fb90}.

Pulsar timing is a complimentary approach to gravitational wave detection that uses milli-second pulsars as a natural galactic scale gravitational wave detector~\cite{Hobbs:2017oam}. Possible sources in the nano-Hertz frequency range probed by pulsar timing arrays include slowly inspiraling supermassive black hole binaries \citep[e.g.,][]{bbr80}, cosmic string networks \citep[e.g.,][]{Damour:2004kw}, and processes in the very early Universe, such as inflation or phase transitions \citep[e.g.,][]{g05}. Over fifty millisecond pulsars, widely distributed across the sky, are now monitored as part of the worldwide pulsar timing effort \citep{Verbiest:2016vem}. Each Earth-pulsar line of sight provides a different projection of the gravitational wave polarization pattern, offering a distinct advantage over existing ground-based interferometers which provide very few independent projections. Moreover, pulsar timing arrays operate in the limit where the wavelengths are much shorter than the light path, while ground-based interferometers operate in the long-wavelength limit. The response to longitudinal polarizations is significantly enhanced relative to the transverse modes for pulsar timing~\cite{2008ApJ...685.1304L, daSilvaAlves:2011fp, Chamberlin:2011ev, grt15}, but not for ground-based interferometers. 

Here we show that existing results from pulsar timing arrays can be used to set stringent limits on the energy density in alternative polarization modes for both astrophysical and cosmological stochastic backgrounds.  We derive expressions for the power spectra of gravitational waves from a population of supermassive black hole binaries. The power spectra for the scalar and vector modes include an additional dipole contribution, which impacts both the generation of the waves and the orbital decay. The measured power spectrum is further modified by the different response functions for scalar, vector and tensor modes. Published upper limits on the power spectrum of pulsar timing residuals can be converted into upper limits on the amplitude of each polarization mode. Note that in our analysis we consider all modes simultaneously. The bounds are particularly strong for the scalar longitudinal and vector longitudinal modes due to the enhanced response to these polarization states~\cite{2008ApJ...685.1304L}. In principle our upper limits can be translated into bounds on the coupling constants for particular alternative theories, but this requires assumptions be made about the merger rate of black holes. For some theories of gravity our results provide no constraints: for example, black hole binaries in Brans-Dicke gravity radiate no differently than in general relativity due to extended no-hair theorems~\cite{Hawking:1972qk}.  For a large class of theories, however, our results do provide constraints because black holes acquire either scalar or vector hair, and thus, emit dipole radiation when in a binary~\cite{lrr-2013-9,Yagi:2013ava,Yagi:2011xp,Yagi:2012vf}.  

\noindent {\em Computing the timing residuals:} 
A stochastic gravitational wave background will produce correlated perturbations in the pulse arrival times measured for pulsars $a$ and $b$ with cross-spectral density
\begin{equation}
S_{ab}(f) = \sum_{I} \Gamma^I_{ab}(f) \, \frac{ h_{c,I}^2(f) }{8 \pi^2 f^3} \, ,
\end{equation}
where $h_{c,I}$ is the characteristic amplitude of the $I^{\rm th}$ polarization state, and $\Gamma_{ab}$ is a geometrical factor that describes the correlation between the pulsars. 
Astrophysical limits are traditionally reported in terms of the amplitude of the characteristic strain at a period of one year: ${\cal A} = h_c(f = {\rm yr}^{-1})$, while cosmological limits are traditionally reported in terms of the energy density per logarithmic frequency interval, scaled by the closure density, $\Omega(f) h^2 = 2 \pi^2 f^2 h_c^2(f)/(3 H_{100}^2)$. 
The correlation pattern for the transverse tensor states of general relativity was first computed by Hellings and Downs~\cite{Hellings:1983fr}:
\begin{equation}
\Gamma^{\rm TT}_{ab} = \frac{1}{3} \left[ 1 + \delta_{ab} + \kappa_{ab} \left( \ln \kappa_{ab}  - \frac{1}{6} \right) \right]\, ,
\end{equation}
where $\delta_{ab}$ is the Kroneker delta, $\kappa_{ab} = (1-\cos\theta_{ab})/2$, and $\theta_{ab}$ is the angle between the line of sight to pulsars $a$ and $b$. Note that for this study we
use un-normalized correlation functions since it is not possible to normalize the longitudinal modes to unity for $a=b$ as is usually done. Instead we have $\Gamma^{\rm TT}_{aa}=2/3$. The correlation pattern for scalar transverse waves is $\Gamma^{\rm ST}_{ab} = (1+\delta_{ab})/3 - \kappa_{ab}/6$. Closed-form expressions for the scalar and vector longitudinal modes are not available, and have to be computed numerically. Approximate expressions for the autocorrelation terms have been found \citep{Chamberlin:2011ev, grt15}, and are given by $\Gamma^{\rm VL}_{aa} \approx 2\ln(4\pi L_a f) -{14}/{3} + 2\gamma_E$ and
$\Gamma^{\rm SL}_{aa} \approx {\pi^2}fL_a/4  - \ln(4\pi L_a f) +{37}/{24} - \gamma_E$, where $\gamma_E$ is the Euler constant and $L_a$ is the light travel time from pulsar $a$. For typical pulsar timing distances and observation frequencies, the quantity $f L_a$ is of order $10^2$ to $10^4$, which implies that the response to longitudinal modes is much larger than to the transverse modes.

Binary systems of supermassive black holes are expected to be dominant sources of gravitational waves in the pulsar timing band. Some alternative theories of gravity predict that these systems will generate scalar and vector dipole radiation (along with sub-dominant higher moments), in addition to the usual tensor quadrupole radiation. Rather than considering specific theories individually, we can derive a general form for the gravitational wave spectrum, which can then be constrained using existing bounds from pulsar timing observations. Turning these bounds into constraints on the coupling constants for specific theories would require more detailed calculations and assumptions about the number of supermassive black hole binaries. Our derivation is based on the analysis in Refs.~\cite{Chatziioannou:2012rf, Sampson:2015ada}, and assumes that the binaries are in circular orbits, with the orbital decay dominated by gravitational wave emission.
Neglecting higher moments, the gravitational wave signal from a slowly evolving binary has the generic form~\cite{Chatziioannou:2012rf}
\begin{equation}
{\bf h}(t) = {\bf A}_{D} f(t)^{1/3} + {\bf A}_{Q} f(t)^{2/3}
\end{equation}
where ${\bf A}_D$ and ${\bf A}_Q$ are the polarization tensors for the dipole and quadrupole modes, scaled by masses, distances, and coupling constants. Note that pulsar timing is generally more sensitive to the dipole terms, since the binaries are well separated, while ground-based detectors are more sensitive to the quadrupole terms since the binaries are close to merger.
We assume that any modifications to the
conservative dynamics are sub-dominant compared to the modifications to the radiative sector, so that to leading order the frequency is related to the orbital separation by Kepler's law $f(t)\sim r(t)^{-3/2}$.
The energy flux $dE/dt$ in general relativity is computed from $\dot{h}^2\sim f^2 h^2$ and an integration over a sphere surrounding the source. In alternative theories the energy flux will also include energy carried by any of the additional fields that must exist in the non-GR theory, but the frequency dependence will be the same since it follows from the multipole decomposition.  Thus we have
\begin{equation}
\frac{d E}{dt} = B_{D} f^{8/3} + B_{Q} f^{10/3}
\end{equation}
where $B_{D}$ and $B_Q$ are related to scalars formed from the squares of ${\bf A}_D$ and ${\bf A}_Q$ integrated over the sphere, along with additional factors that come from the scalar and vector degrees of freedom. Combining this with the Newtonian expression for the binding energy $E=-G m\mu/r \sim f^{2/3}$, we have $dE/df \sim f^{-1/3}$, and
\begin{equation}
\frac{d f}{dt} =  \frac{d E/dt}{dE/df} = C_{D} f^{9/3} + C_{Q} f^{11/3}
\end{equation}
Combining the expressions for ${\bf h}(t)$ and $df/dt$ according to the formalism in Ref.~\cite{Sampson:2015ada} yields
\begin{eqnarray}
&& S_{ab}(f) = \left( \frac{1+\kappa^2}{1 + \kappa^2 \left(\frac{f}{{\rm yr}^{-1}}\right)^{-2/3}}\right) \left(\Gamma^{\rm TT}_{ab} {\cal A}_{\rm TT}^2\left(\frac{f}{{\rm yr}^{-1}}\right)^{-4/3}  \right. \nonumber \\
&& \left. +  (\Gamma^{\rm ST}_{ab}{\cal A}_{\rm ST}^2 +   \Gamma^{\rm VL}_{ab}{\cal A}_{\rm VL}^2 +  \Gamma^{\rm SL}_{ab}{\cal A}_{\rm SL}^2)\left(\frac{f}{{\rm yr}^{-1}}\right)^{-2} \right)\frac{1}{8 \pi^2 f^3}.
\end{eqnarray}

In addition to signals from binary black holes, the pulsar timing band may contain signals from a network of cosmic strings, or from processes in the early Universe, such as phase transitions or inflation. Computing the gravitational wave signature for each of these sources in a general way for alternative theories of gravity is outside the scope of our current work. One simple case that we can address is inflation, where general considerations imply that the scalar, vector and tensor modes that re-enter the horizon during the radiation dominated epoch will have a nearly scale-invariant spectrum, with
$\Omega h^2$ independent of frequency, and
\begin{equation}\label{cosmo}
S_{ab}(f) = \frac{ 3H_{100}^2 h^2 }{16 \pi^4 f^5} \sum_{I={\rm TT,ST,VL,SL}} \Gamma^I_{ab}(f) \, \Omega_I .
\end{equation}

\noindent {\em Constraints on alternative polarizations:} 
Existing results from pulsar timing studies can be used to place interesting constraints on the energy density of tensor and non-tensor polarization states. While the Parkes Pulsar Timing array
currently has the lowest published upper limit on the tensor amplitude~\cite{Shannon:2013wma}, it is difficult to map those limits to constraints on other polarization states that have different spectra.
Instead, we chose to use the Bayesian per-frequency upper limits on $h_c(f)$ derived by the NANOGrav collaboration~\cite{Arzoumanian:2015liz}, from which we can derive a likelihood function for $S_d(f)=S_{aa}(f)$. Since the NANOGrav bounds are for tensor modes, we have the mapping $S_{d}(f) = h_c(f)^2/(12 \pi^2 f^3)$.  Following Ref.~\cite{Middleton:2017nbg}, we model the per-frequency posterior distributions for $h_c$ with Fermi functions:
\begin{equation}
p(h_c) = \frac{1}{\sigma \ln\left(e^{\frac{h_*}{\sigma}}+1\right)\left(1+e^{\frac{(h_c-h_*)}{\sigma}}\right)}
\end{equation}
with $\sigma \approx h_*/2$ and the turn-over point $h_*$ is related to the quoted 95\% upper limits $h_{95}$ by
\begin{equation}
h_{95} = h_* - \sigma \ln\left( \left( e^{h_*/\sigma}+1\right)^{0.95} -1\right).
\end{equation}
The posterior distributions for $h_c(f)$ define a posterior distribution for $S_{d}(f)$, which we then use to define a likelihood for the model parameters ${\cal A}_I,\kappa$ or $\Omega_I$ from the product $\prod_f p(S_{aa}(f))$. Applying this procedure to a purely tensor theory yields the 95\% upper limits ${\cal A} < 1.9 \times 10^{-15}$ and $\Omega h^2 < 7.4 \times 10^{-10}$, which are in reasonable agreement with the directly computed upper limits~\cite{Arzoumanian:2015liz}, ${\cal A} < 1.5 \times 10^{-15}$ and $\Omega h^2 < 4.2 \times 10^{-10}$. The discrepancies are likely due to imperfections in our fit to the $h_c(f)$ posterior distributions, and differences in the covariances between red noise model and the signal model in the per-frequency-bin versus full spectrum analyses.  The bounds we derive will provide conservative upper limits on the alternative polarization states.

One additional caveat that pertains to using the previously derived bounds on $S_{aa}(f)$ to constrain alternative polarization states is that the original analysis combines the limits derived from multiple different pulsars, each at a different distances from Earth. Ideally, we would use the per-pulsar bounds on $S_{aa}(f)$ and factor in the different distances to each pulsar, which enter into the response function for the longitudinal modes. This information, however, is not publicly available, and since the best timed pulsars are all at roughly the same distance from Earth, we simply assume that all the pulsars are at a distance of $1\pm 0.2$ Kpc from Earth, and marginalize over the uncertainty.

 \begin{figure}[htp]
\includegraphics[clip=true,angle=0,width=0.46\textwidth]{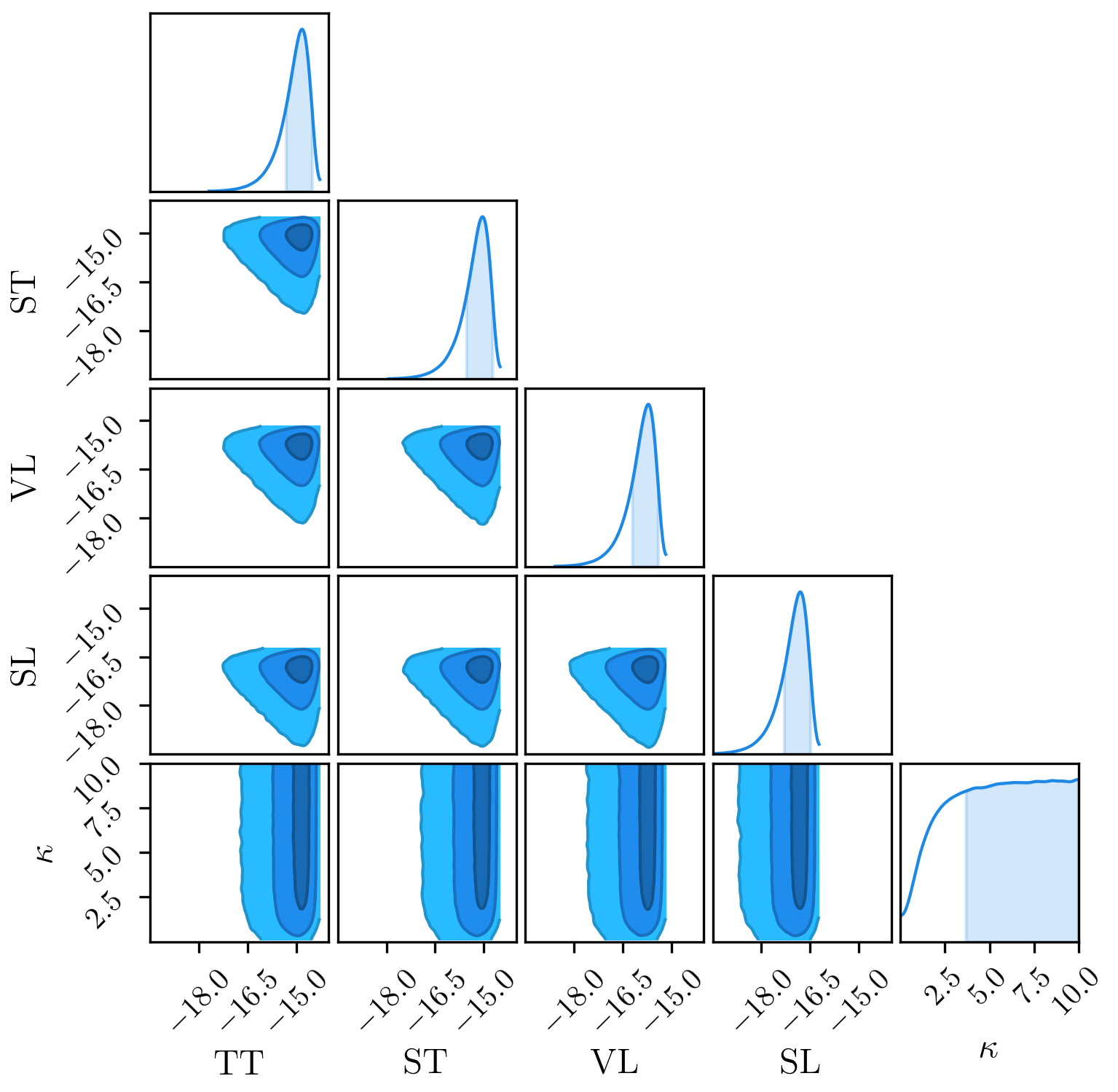} 
\caption{\label{fig:astro} Slices through the posterior distribution for the astrophysical amplitudes ${\cal A}_{\rm TT}$, ${\cal A}_{\rm ST}$, ${\cal A}_{\rm VL}$, ${\cal A}_{\rm SL}$  and the decay parameter $\kappa$. }
\end{figure}

\begin{figure}[htp]
\includegraphics[clip=true,angle=0,width=0.46\textwidth]{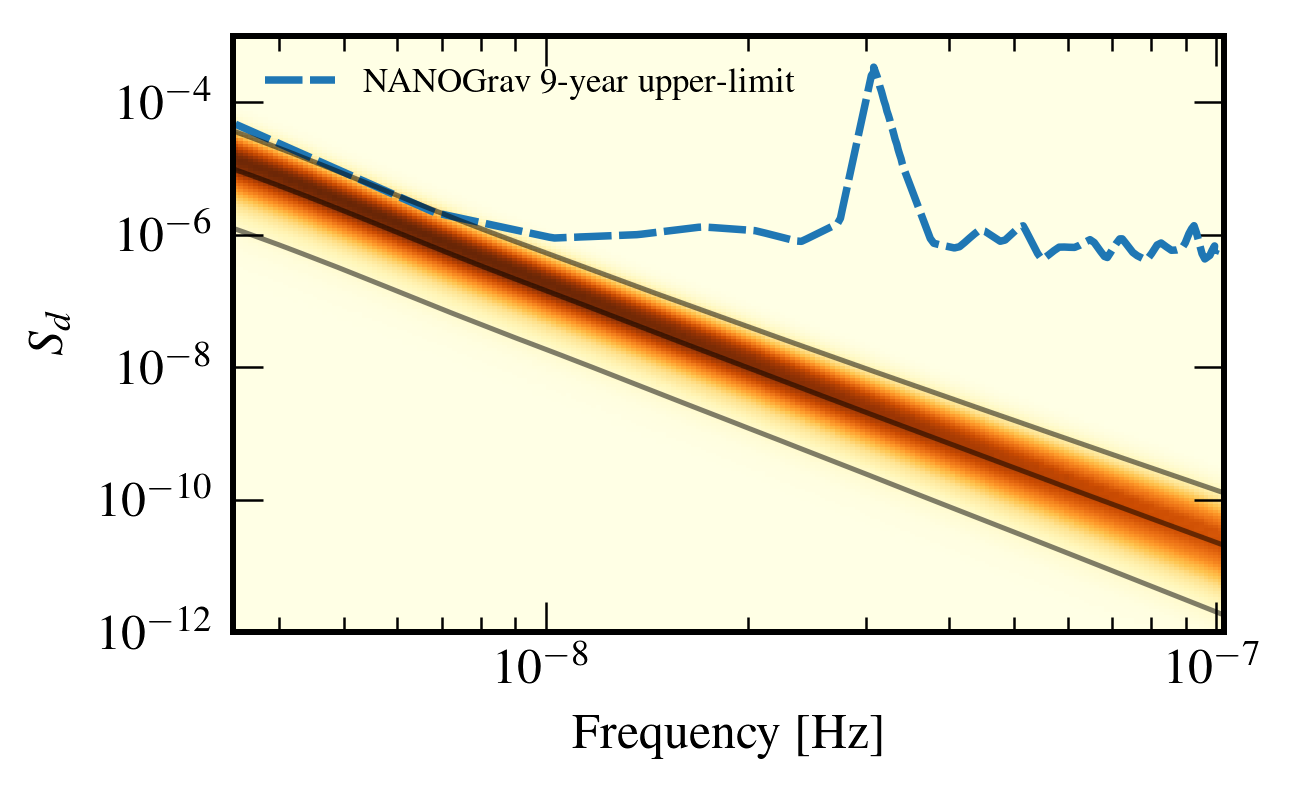} 
\caption{\label{fig:bg} Posterior distribution for the combined spectrum $S_{d}(f)$ of the astrophysical model (shaded region) compared to the NANOGrav 9-year upper limits (dashed line).}
\end{figure}

Results for the amplitudes of the astrophysical signal are shown in Fig.~\ref{fig:astro} assuming uniform priors on the amplitudes $0 \leq {\cal A} < 10^{-13}$ and decay parameter $0\leq \kappa < 10$. Note that values of $\kappa > 10$ produce spectra in the PTA band that are identical to those with $\kappa =10$, hence our choice of upper bound on the prior range. The 95\% upper limits on the amplitudes are ${\cal A}_{\rm TT} < 3.3\times 10^{-15}$, ${\cal A}_{\rm ST} < 1.9 \times 10^{-15}$, ${\cal A}_{\rm VL} < 4.1\times 10^{-16}$ and ${\cal A}_{\rm SL} < 3.7\times 10^{-17}$. The posterior distribution for the spectrum $S_d(f)$ is plotted against the NANOGrav 9-year upper limits~\cite{Arzoumanian:2015liz} in Fig.~\ref{fig:bg}.
Repeating the analysis for the cosmological model in Eq.~(\ref{cosmo}), we find 95\% upper limits on the energy densities of $\Omega_{\rm TT+ST}h^2  < 7.7  \times 10^{-10}$, $\Omega_{\rm VL}h^2  < 3.5  \times 10^{-11}$ and $\Omega_{\rm SL}h^2  < 3.2  \times 10^{-13}$. Note that we can only constrain the sum of energy density in the tensor and scalar transverse modes since they produce residuals with identical frequency dependence.

\begin{figure}[htp]
\includegraphics[clip=true,angle=0,width=0.46\textwidth]{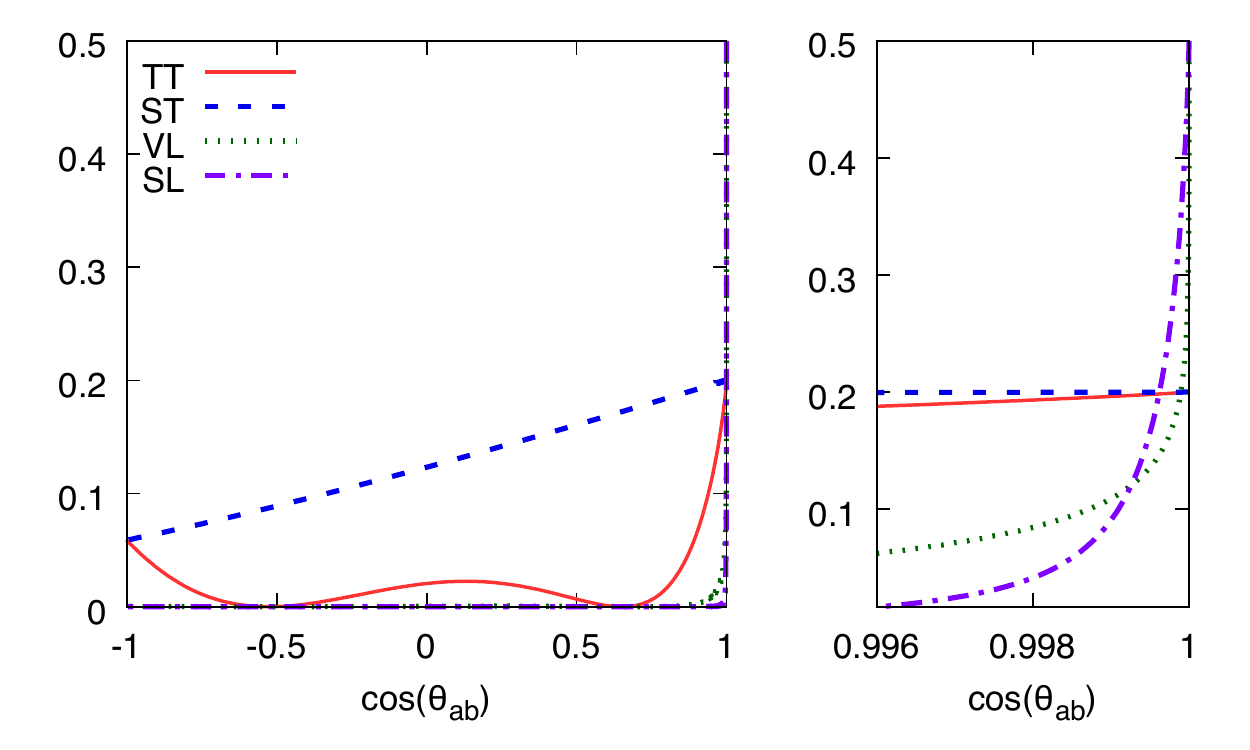} 
\caption{\label{fig:sum} The contribution to the signal-to-noise ratio squared as a function of the cosine of the angle between the pulsars for the various polarization states.
The panel on the right highlights the small region near zero angular separation.}
\end{figure}

\noindent {\em Detecting alternative polarization states:}
The unique angular correlation patterns imprinted by gravitational waves should allow us to distinguish between a stochastic gravitational wave background and the myriad sources of noise that impact pulsar timing. What we discovered when analyzing simulated signals came as a surprise: we found that the longitudinal modes made it very difficult to detect any correlation pattern, even in the zero noise limit. In effect, the longitudinal signals behave as noise. The signal we are looking for is the cross-spectrum of the timing residuals $S_{ab}(f) = {\rm E}[r_a(f)r_b(f)]$, which have variance $\sigma^2_{ab} =
{\rm E}[r_a(f)r_b(f)r_a(f)r_b(f)]-{\rm E}[r_a(f)r_b(f)]^2 = S_{aa}(f) S_{bb}(f) + S^2_{ab}(f)$. The variance of the longitudinal modes is very large due to the $fL$ dependence in the auto-covariance terms.  We can quantify this effect by computing the signal-to-noise ratio of the $X_B$ correlation statistic~\cite{Rosado:2015epa}. We consider the observation-noise-free limit, for if the signal cannot be detected without noise, it will not be detectable with noise. In the zero observation-noise limit, the signal-to-noise ratio squared of the $X_B$ statistic for the $I^{\rm th}$ polarization state is 
\begin{equation}
{\rm SNR}^2_B = \left( 2 \sum_{f} \sum_a^{N_p} \sum_{b > a}^{N_p} \frac{ {{\Gamma}_{ab}^I}^2(f) }{\Gamma_{aa}^I(f) \Gamma_{bb}^I(f) +{{ \Gamma}_{ab}^I}^2(f)} \right) \, .
\end{equation}
The Bayesian evidence for a correlation being present scales as ${\rm SNR}^2_B$.
The angular dependence of the summand for each polarization state is shown in Fig.~\ref{fig:sum} for $f=10^{-8}$ Hz and $L=1$ Kpc. We see that the longitudinal modes accumulate most of their signal-to-noise ratio from pulsars with very small angular separations.

\begin{figure}[htp]
\includegraphics[clip=true,angle=0,width=0.46\textwidth]{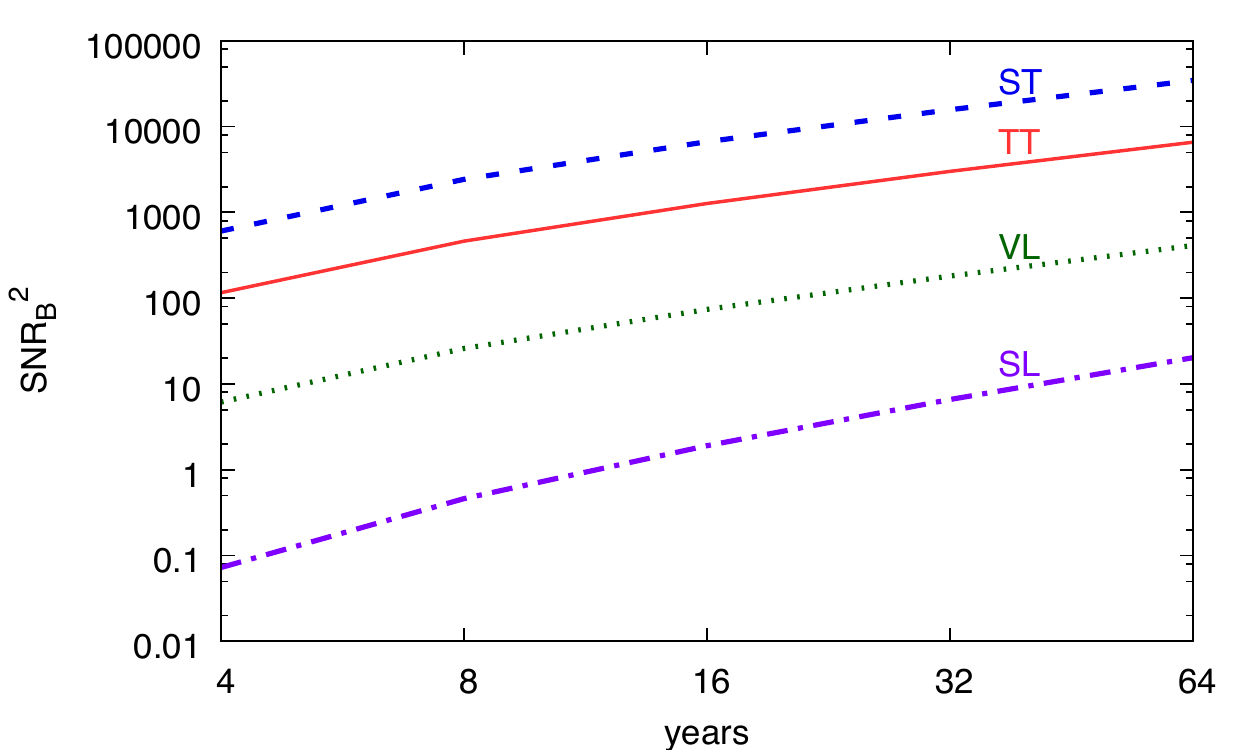} 
\caption{\label{fig:snr} The growth in the signal-to-noise-ratio squared with time for each polarization mode, assuming that the 46 pulsars of the International Pulsar Timing Array are in the signal-dominated regime.}
\end{figure}

The relative detectability of the various polarizations can be illustrated by considering the 46 pulsars from International Pulsar Timing Array~\cite{Verbiest:2016vem}.  Figure~\ref{fig:snr} shows the scaling of the signal-to-noise-ratio-squared as a function of the observation time $T$ under the assumption that the signal dominates the noise for $3/T \leq f \leq {\rm yr}^{-1}$.  The vector longitudinal modes are a factor of ten harder to detect than the tensor modes, while the scalar longitudinal mode is a factor of a thousand harder to detect. With enough pulsars and a long enough observation time it will be possible to separate the scalar, vector and tensor modes, but the observational challenge is much greater than originally thought~\cite{2008ApJ...685.1304L}. The difference in our conclusions can be traced to the original study using a detection statistic that neglects the auto-correlation terms.

{\em Summary:} 
We have derived the first pulsar timing bounds on the amplitude of scalar and vector stochastic gravitational wave backgrounds for both astrophysical and cosmological sources. We have also pointed out that the ``self-noise'' produce by the strong response to longitudinal modes will make detecting alternative polarization states from a stochastic background very challenging. We hypothesize that observations of bright resolvable systems may provide the best opportunity to probe alternative polarization states using pulsar timing, since there the autocorrelation terms will contribute to the signal, not the noise.

\section*{Acknowledgments}
We appreciate Laura Sampson's help with generating simulated data sets. We thank Xavier Siemens for suggesting that we extend our study to cover cosmological backgrounds, and we thank Justin Ellis for providing the posterior samples for the characteristic strain that were used to calibrate our likelihood model. LO, NJC and SRT appreciate the support of the NSF Physics Frontiers Center Award PFC-1430284. The research was partially carried out at the Jet Propulsion Laboratory, California Institute of Technology, under a contract with the National Aeronautics and Space Administration. NY also acknowledges support from the NSF CAREER grant PHY-1250636 and NASA grants NNX16AB98G and 80NSSC17M0041.
\bibliography{refs}

\end{document}